\documentstyle[preprint,prl,aps]{revtex}
\newcommand{\beq}{\begin{equation}}
\newcommand{\eeq}{\end{equation}}
\newcommand{\bea}{\begin{eqnarray}}
\newcommand{\eea}{\end{eqnarray}}
\newcommand{\be}{\begin{enumerate}}
\newcommand{\ee}{\end{enumerate}}
\newcommand{\bi}{\begin{itemize}}
\newcommand{\ei}{\end{itemize}}
\newcommand{\ba}{\begin{array}}
\newcommand{\ea}{\end{array}}
\newcommand{\bc}{\begin{center}}
\newcommand{\ec}{\end{center}}
\newcommand{\bt}{\begin{tabular}}
\newcommand{\et}{\end{tabular}}
\newcommand{\half}{\textstyle {1\over2} \displaystyle}    
\newcommand{\Dslash}{{\hbox{D}\kern-0.6em\raise0.15ex\hbox{/}}} 

\begin{document}


\draft

\title{Random Ising Spins in Two Dimensions - A Flat Space Realization
of the KPZ Exponents}

\author{ Marco Veki\'c}
\address{
Institute for Theoretical Physics,
University of California
Santa Barbara, CA 93106
}

\author{Shao Liu }
\address{
Department of Physics,
University of California
Irvine, CA 92717
}

\author{ Herbert W. Hamber}
\address{Theory Division,
CERN,
CH-1211 Gen\`{e}ve 23, Switzerland}

\date{\today}
\maketitle
\begin{abstract}
A model describing Ising spins with short range interactions moving
randomly in a plane is considered. In the presence of a hard core repulsion,
which prevents the Ising spins from overlapping, the model is analogous
to a dynamically triangulated Ising model with spins constrained
to move on a flat surface.
As a function of coupling strength and hard core repulsion
the model exhibits multicritical behavior, with first and second order
transition lines terminating at a tricritical point.
The thermal and magnetic exponents computed at the tricritical point
are consistent with the KPZ values associated with Ising spins,
and with the exact two-matrix model solution
of the random
Ising model, introduced previously to describe the effects of
fluctuating geometries.
\end{abstract}




\newpage

\narrowtext

\section{Introduction}

Following the exact solution of the Ising model on a random surface by
matrix model methods \cite{kazakov}, there has been growing interest in
the
properties of random Ising spins coupled to two-dimensional gravity.
More recently, work based on both series expansions \cite{bh,cicuta}
and numerical simulations \cite{dtrsis,phicube} has verified and
extended
the original results.
It is characteristic of these Ising models that the spins are allowed
to move at random on a discretized version of a fluid surface.
In a specific implementation of the model, Ising spins are placed at
the vertices of a lattice built out of equilateral triangles, and the
lattice geometry is then allowed to fluctuate by varying the local
coordination number through a ``link flip'' operation which varies the
local connectivity \cite{dtrsis}.
Remarkably, the same critical exponents have also been found using
consistency
conditions derived from conformal field
theory for central charge $c=\half$ \cite{kpz}, which should again apply
to
Ising spins.
It is generally believed that the new values for the Ising critical
exponents are due to the random fluctuations of the surface
in which the spins are embedded,
and therefore intimately tied to the intrinsic fractal properties of
fluctuating geometries.
It came therefore as a surprise that non-random Ising spins, placed
on a randomly fluctuating geometry but with fixed spin coordination
number, exhibited the same critical behavior as in flat space, without
any observed ``gravitational'' shift of the exponents \cite{gh}.

The natural question is then to what extent the values of the critical
Ising exponents found by KPZ for $c=\half$ and in the matrix model solution
($\alpha = -1$, $\beta = 1/2 $, $\gamma = 2$,
$\eta = 2/3 $, $\nu = 3/2$ \cite{kazakov,kpz})
are due to the
{\it annealed} randomness of the lattice, and to what extent they are
due
to the physical presence of a fluctuating background metric.
The most straightforward
way to answer this question is to investigate the critical properties
of annealed random Ising spins, with interactions designed to mimic
as closely as possible the dynamical triangulation model,
but placed in flat two-dimensional space.
It is well known that for a {\it quenched}
random lattice the critical
exponents are the same as on a regular lattice \cite{rlattis},
as expected on the basis of universality, even though in two dimensions
the Harris criterion (which applies to quenched
impurities only) does not give a clear prediction, since the specific
heat exponent vanishes, $\alpha=0$, for Onsager's solution.

In this paper we present some detailed results concerning the exponents
of such a model in order to complete the discussion presented in a recent
publication \cite{Hamber}.

\vskip 10pt
\section{ Formulation of the Model}

In a square $d$-dimensional box of sides $L$ with periodic boundary
conditions we
introduce a set of $N=L^d$ Ising spins $S_i = \pm 1$ with
coordinates $x_i^a$, $i=1...N$, $a=1...d$, and average density
$\rho = N / L^d = 1 $.
Both the spins and the coordinates will be considered as dynamical
variables in this model.
Interactions between the spins are determined by
\beq
I[x,S] \; = \; - \sum_{ i < j } J_{ij} ( x_i , x_j ) \; W_{ij} \; S_i S_j
- h \sum_i W_i \; S_i \;\;\;\; ,
\label{eq:ac}
\eeq
with ferromagnetic coupling
\beq
  J_{ij}(x_i,x_j) = \left \{ \begin{array}{ll}
  0      & \mbox {if $| x_i - x_j | > R$     } \\
  J      & \mbox {if $r < | x_i - x_j | < R$ }
\end{array}
\right .
\;\;\;\; ,
\label{eq:j}
\eeq
and infinite energy for $| x_i - x_j | < r$,
giving therefore a hard core repulsion radius equal to $r/2$.
As will be discussed further below, the hard core repulsive
interaction is necessary
for obtaining a non-trivial phase diagram, and mimics the
interaction found in the dynamical triangulation model, where the
minimum distance between any two spins is restricted to be one lattice
spacing.
For $r \rightarrow 0 $, $J_{ij} = J [ 1 - \theta ( | x_i - x_j | - R ) ] $.

The weights $W_{ij}$ and $W_i$ appearing in Eq.~(\ref{eq:ac}) could in
principle contain
geometric factors associated with the random lattice subtended by the
points, and involve quantities such as the areas of the triangles
associated with the vertices, as well as the lengths of the edges
connecting the sites. In the following we will consider only the simplest
case of unit weights, $W_{ij}=W_i=1$. On the basis of universality of
critical behavior one would expect that the results should not be too
sensitive to such a specific choice, which only alters the short
distance details of the model, and should not affect the exponents.

The full partition function for coordinates and spins is then written as
\beq
Z = \prod_{i=1}^N \sum_{S_i = \pm 1} \bigl ( \prod_{a=1}^d \int_0^L dx_i^a
\bigr ) \; \exp ( - I[x,S] ) \;\;\;\; .
\label{eq:z}
\eeq
In the following we will only consider the two-dimensional case, $d=2$,
for which specific predictions are available from the work of KPZ and
the matrix model solution.

It should be clear that if the interaction range $R$ is of order one, then,
for sufficiently large hard core repulsion,
$r \rightarrow \sqrt{5}/2 < R $,
the spins will tend to lock in into an almost regular triangular
lattice. As will be shown below, in practice this crossover happens
already for quite small values of $r$.
The critical behavior is then the one expected for the
regular Ising model in two dimensions, namely a continuous
second order phase transition with the Onsager exponents.
Indeed for the Ising model on a triangular lattice it is known that
$J_c= \half \sqrt{3} \ln 3 = 0.9514...$.
On the other hand
if the hard core repulsion is very small, then for sufficiently
low temperatures the spins will tend to form tight ordered clusters,
in which each spin interacts with a large number of neighbors.
As will be shown below, this clustering transition is rather sudden and
strongly first order. Furthermore, where the two transition lines meet
inside the phase diagram one would expect to find a tricritical point.

In order to investigate these issues further, we
have chosen to study the above system by numerical simulation, with
both the spins and the coordinates updated by a standard Monte Carlo
method. The computation of thermodynamic averages is quite time
consuming in this model, since any spin can in principle interact with any
other spin as long as they get sufficiently close together.
As a consequence, a sweep through the lattice requires a number of order
$N^2$ operations, which makes it increasingly difficult to study
larger and larger lattices. In order to extend our study to even larger
lattices, we have applied a binning procedure in such a way that the
time for the
updating of a given configuration grows as $zN$, where $z$ is the
average coordination number of the lattice, instead of $N^2$. This binning
procedure consists of dividing the system in cells of unit length, and
keeping track of the spins in each cell. Since all the moves are local, and
spins can only move from a given cell to the neighboring ones,
we only need to consider the spins in a given cell and
its neighbors at each
updating step. This procedure is very effective when the average
coordination
number is relatively small ($J<J_c$ and $r$ large), however,
if $z\sim N$ the updating time grows again as $N^2$.

In this paper we will not address in detail the problem of critical
slowing down, however an additional
possibility for the future could be to implement some sort of cluster updating
algorithm \cite{cluster}.
On the other hand, we should add that
we have not found
any anomalous behavior as far as the autocorrelation times are
concerned, which remain quite comparable to the pure Ising case.

There are a number of local averages and fluctuations which can be
determined and used to compute the critical exponents.
In the course of the simulation the spontaneous magnetization per spin
\beq
M = {1 \over N} {\partial \over \partial h} \ln Z |_{h=0} =
{1 \over N} < | \sum_i S_i | > \;\;\; ,
\label{eq:m}
\eeq
was measured (here the averages involve both the $x$ and $S$
variables, $< \; > \equiv < \; >_{x,S}$), as well as
the zero field susceptibility
\beq
\chi = {1 \over N} {\partial^2 \over \partial h^2} \ln Z |_{h=0} =
{1 \over N} < \sum_{ij} S_i S_j > - {1 \over N} < | \sum_{i} S_i | >^2 \; .
\label{eq:chi}
\eeq
It is customary to use the absolute value on the r.h.s., since
on a finite lattice the spontaneous magnetization, defined without
the absolute value, vanishes identically even at low temperatures.
In addition, in order to determine the latent heat and the specific
heat exponent, we have computed the average Ising energy
per spin defined here as
\beq
E \; = \; - {1 \over N} {\partial \over \partial J} \ln Z |_{h=0} =
- {1 \over J N} < \sum_{ i < j } J_{ij} ( x_i , x_j ) \; W_{ij} \; S_i
S_j > \;\;\;\; ,
\label{eq:e}
\eeq
and its fluctuation,
\beq
C \; = \; {1 \over N} {\partial^2 \over \partial J^2} \ln Z |_{h=0}
\;\;\;\; .
\label{eq:c}
\eeq
Some additional quantities we have used in the course of this work
will be defined later.

\vskip 10pt
\section{ Results and Analysis }

In the simulations we have investigated lattice sizes varying
from $5^2= 25$ sites to
$30^2 = 900$ sites.  The length of our runs varies in the critical
region ($J \sim J_c $) between 2M sweeps on the smaller lattices
and 200k sweeps on the largest lattices. A standard binning procedure
then leads to the errors reported in the figures.

As it stands, the model contains three coupling parameters,
the ferromagnetic coupling $J$, the interaction
range $R$ and the hard core repulsion parameter $r$. We have fixed
$R=1$;
comparable choices should not change the universality class.
As we alluded previously, for small $r$ we find that the system
undergoes a sharp first order transition, between the disordered
phase and a phase in which all spins form a few very tight magnetized
clusters, in which the number of neighbors is of the order $N$.
These clusters persist even for larger values of the hard core
repulsion, $r$,
but the number of interacting neighbors does not become as large as $N$
in this case.

In Figs.\ \ref{config2} and\ \ref{config1} we show the existence of
these clusters when the hard core repulsion is as large as $r=0.4$. In
Fig.\ \ref{config2} we observe ferromagnetic order in small
domains even though we are below $J_c$. On the other hand, in Fig.\
\ref{config1}, where we are above $J_c$, the system has practically
clustered into a single ferromagnetic domain.
For sufficiently large $r$, the transition is
Ising-like, between ordered and disordered, almost regular, Ising
lattices (for our choice of range $R$, the transition appears to
be very close to regular Ising-like for $r \approx 0.6$ and larger,
see below). In
Fig.\ \ref{config3} we show a particular configuration for $r=0.98$
where the regular, almost triangular, lattice is clearly visible. In this
case the average
coordination number $z$ is very close to $3$, as expected for a regular
triangular lattice. In Fig.\ \ref{neighbors} we show the average number
of neighbors $z$ for several values of $r$ on a system with $N=144$ spins.
We find that for small values of $r$ the coordination number increases
very rapidly as we approach the critical point. On the other hand,
for intermediate choices of $r$, $z$ saturates
to a smaller value. When $r=0.6$, the coordination number
saturates to a value of $z= 3.1$, which is
already very close to the value on a regular triangular lattice ($z=3$).

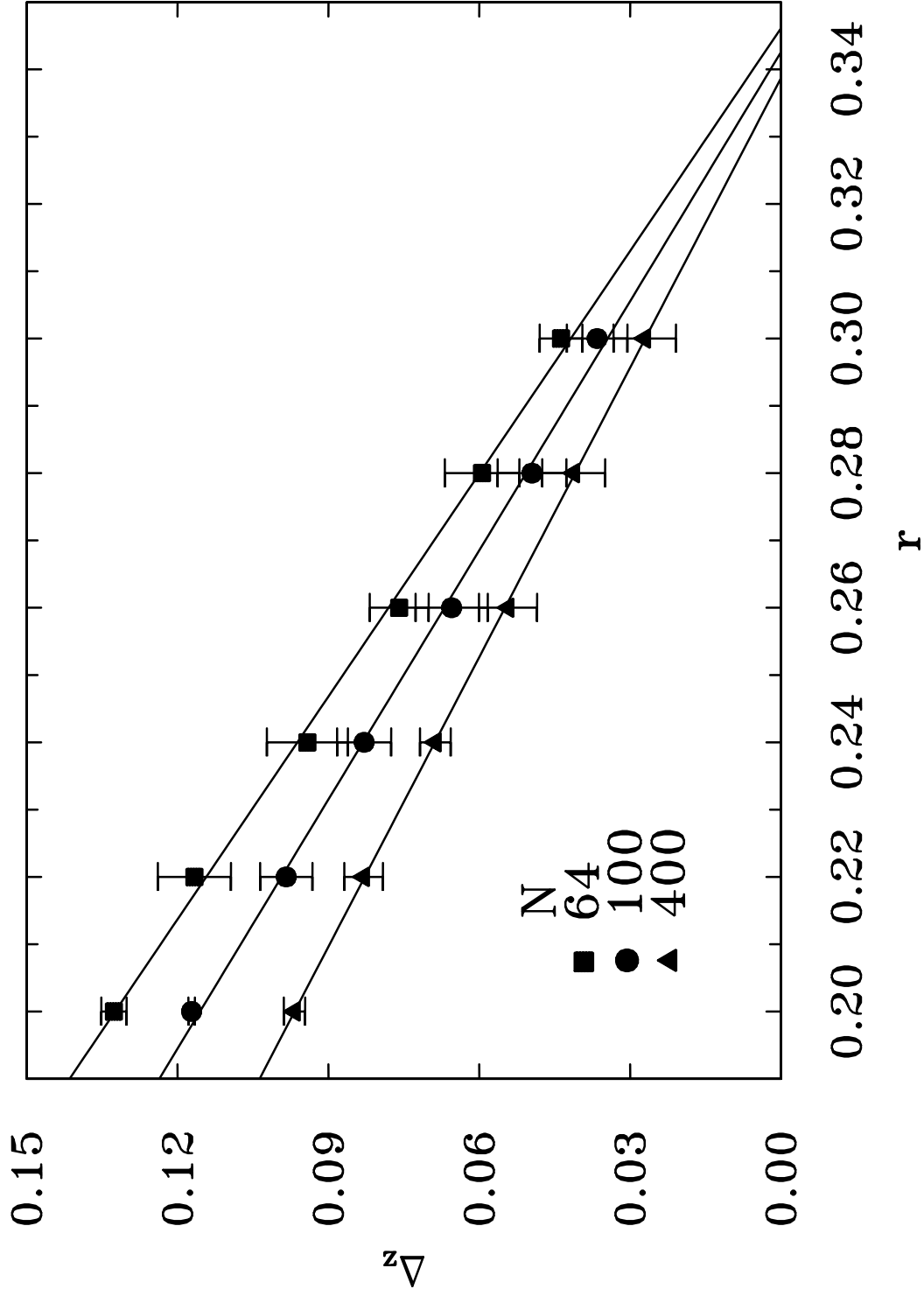
\begin{figure}
\caption{
The latent heat per bond $\Delta_z$ along the first order transition line,
plotted against
the hard core repulsion parameter $r$.
The tricritical point is located where the latent heat vanishes.
}
\label{latent}
\end{figure}

In Fig.\ \ref{energy} we plot the average energy per bond $E_z$
as a function of
$J$
for several choices of the hard core repulsion $r$. The jump
discontinuity, which is visible for small hard core repulsion $r$,
indicates the existence of a first order transition. For larger
values of $r$, the discontinuity is reduced and eventually vanishes.
A determination of the discontinuity in the average energy of
Fig.\ \ref{energy}  at the
critical coupling $J_c$ shows that it gradually decreases as $r$ is
increased from zero.
Fig.\ \ref{latent} shows a plot of the latent heat per bond $\Delta_z$
versus $r$ at the
transition point $J_c$.
In general we do not expect the latent heat to vanish linearly at
the endpoint, but our results seem to indicate a behavior quite
close to linear. From
the data we estimate that the latent heat vanishes
at $r=0.344(7)$, thus signaling the presence of a tricritical point at
the end of the first order transition line. Beyond this point, the
transition stays second order, as will be discussed further below.
The phase transition line is shown in Fig.\ \ref{Isingdiag};
for $r=0$ we found on the
largest lattices $J_c=0.19(2)$, while for $r=0.98$ we found
$J_c=0.93(3)$.

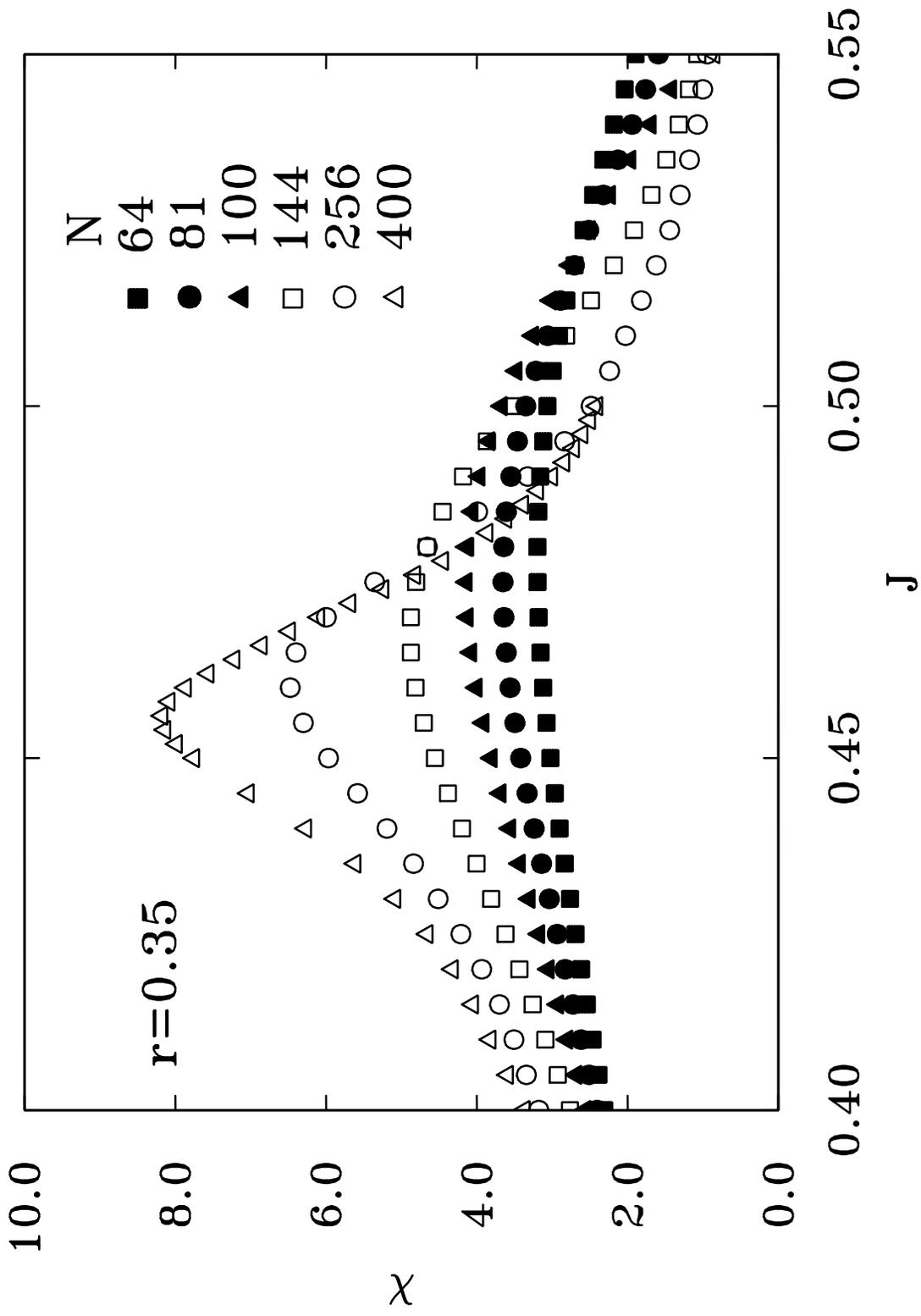
\begin{figure}
\caption{
The magnetic susceptibility $\chi$ versus $J$
for fixed hard core repulsion parameter $r=0.35$ and different system
sizes.
}
\label{suscN}
\end{figure}
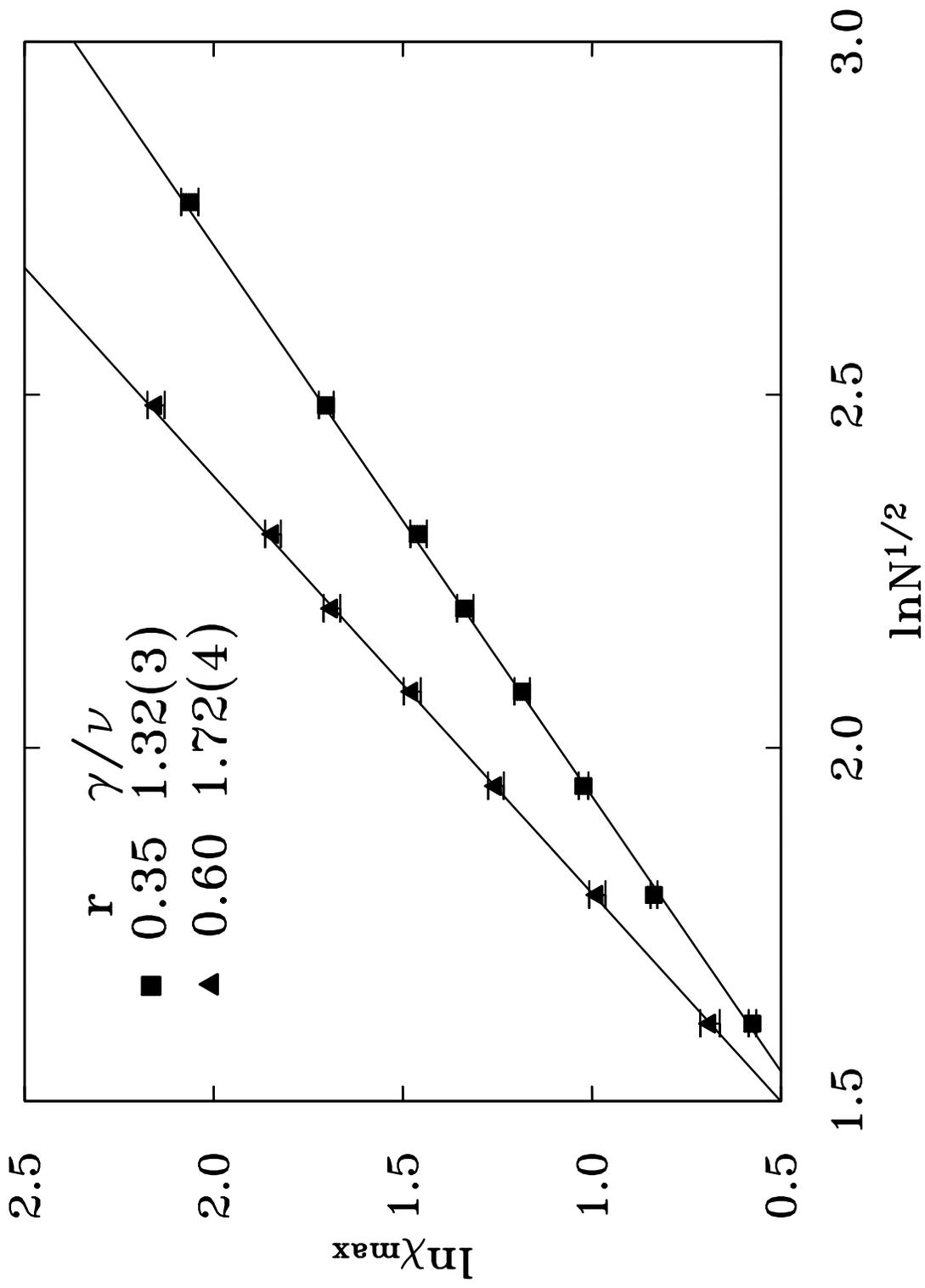
\begin{figure}
\caption{
The peak in the magnetic susceptibility $\chi_{\mbox{\tiny max}}$
versus the number of
Ising spins $N$ for choices of the
hard core repulsion parameter corresponding to $r=0.35$ and $r=0.6$.
}
\label{susc}
\end{figure}

In Fig.\ \ref{suscN} we plot the spin susceptibility as a function of
$J$ for
several system sizes near the tricritical point,
showing a growth of the peak with system
size.
To determine the critical exponents, we will resort to a finite-size
scaling analysis. In the following we will be mostly concerned with
the values for the critical exponents in the vicinity of the
tricritical point.
In the case of the spin susceptibility, from finite-size scaling,
we expect
a scaling form of the type
\beq
\chi (N,J) = N^{\gamma / 2 \nu } \; \bar \chi ( N^{1 / 2 \nu} |J -J_c|)
\;\;\;\; .
\eeq
To recover the correct infinite volume result one needs
$ \bar \chi ( x ) \sim x^{-\gamma} $ for large arguments.
Thus, in particular the peak in $\chi$ should scale like
$ N^{ \gamma / 2 \nu } $ for sufficiently large $N$. In Fig.\ \ref{susc}
we show the evolution of the
computed peaks in $\chi$  as a function of $\ln N$.

Despite the fact that the lattices are quite small,
as can be seen from the graph, a linear fit to the data at the
tricritical point is rather good,
with relatively small deviations
from linearity, $\chi^2 / d.o.f. \sim 10^{-4} $.
Using least-squares one can estimate
$ \gamma / \nu $. For $r=0.35$ we find $ \gamma / \nu =1.32(3)$,
which is much smaller
than the exact regular Ising result $ \gamma / \nu = 1.75 $. From
scaling one then obtains the anomalous dimension exponent
$\eta = 2 - \gamma / \nu =0.68(3)$.
To further gauge our errors, we have computed the same exponent for
the regular Ising limit, for $r=0.6$. In this
case we indeed recover the
Onsager value: we find on the same size lattices and using the same
analysis method $ \gamma / \nu = 1.72(4) $.
We also note that the shift in the critical point on a finite lattice
is expected to be determined
by the correlation length exponent $\nu$, namely
$J_c (N) - J_c (\infty) \sim N^{- 1 / 2 \nu }$. This relationship can
be used to estimate $\nu$, but it is not very accurate. From a fit
to the known values of $J_c(N)$ we obtain
a first rough estimate $\nu = 1.3(2)$. A more precise determination
of $\nu$ will be given later.

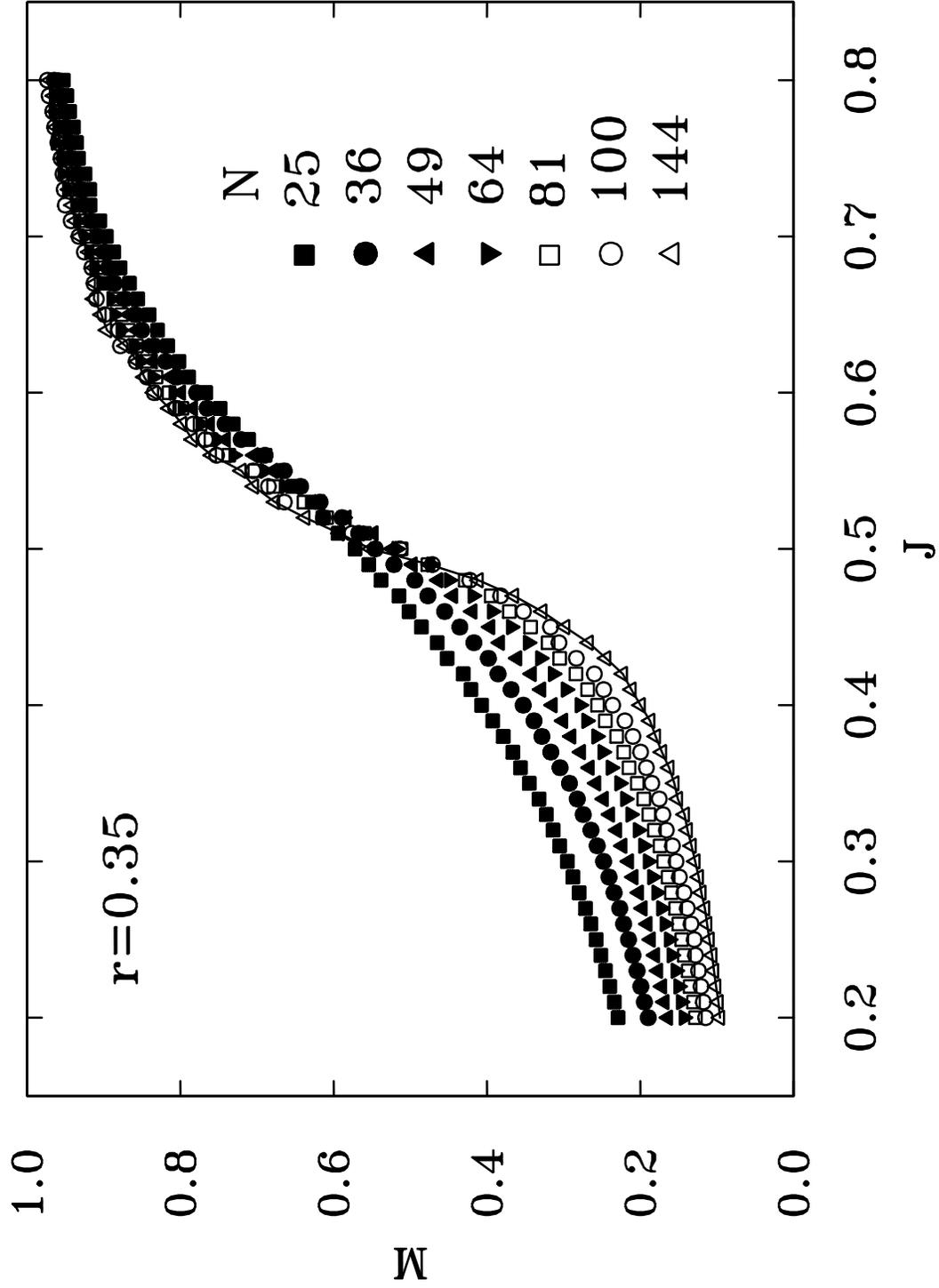
\begin{figure}
\caption{
The magnetization $M$ versus $J$,
for fixed hard core repulsion parameter $r=0.35$ and different system
sizes.
The solid line is a spline through the data for $N=144$.
}
\label{magN}
\end{figure}
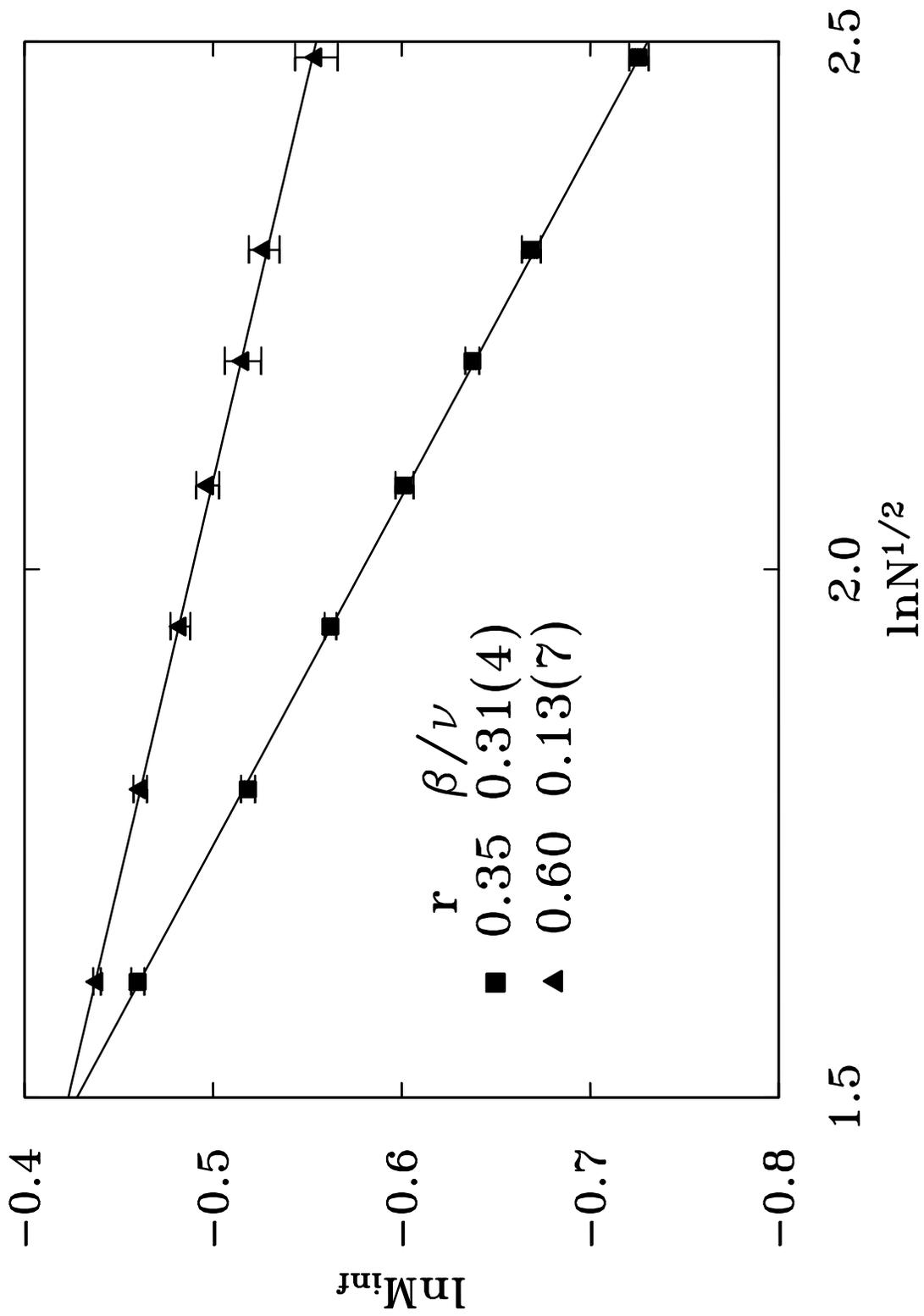
\begin{figure}
\caption{
Finite size scaling of the magnetization at the inflection point
$M_{\mbox{\tiny inf}}$
versus the total number of
Ising spins $N$ for choices of the
hard core repulsion parameter corresponding to $r=0.35$ and $r=0.6$.
}
\label{mag}
\end{figure}

A similar finite-size scaling analysis can be performed for the
magnetization,
which is shown in Fig.\ \ref{magN} for several system sizes.
Close to and above $J_c$ we expect $M \sim (J-J_c)^\beta $.
At the critical point on a finite lattice, as determined from
the peak in the susceptibility (which incidentally is very close to
the inflection point in the magnetization versus $J$),
$M$ should scale to zero as
$M_N(J_c) \sim N^{\beta / 2 \nu }$.
In Fig.\ \ref{mag} we show the magnetization $M$ computed in this way
for different size lattices
At the tricritical point we find $ \beta / \nu =  0.31(4) $, which
again clearly excludes the pure Ising exponent, $ \beta / \nu =  0.125$.
For the pure Ising limit ($r=0.6$) we obtain $ \beta / \nu = 0.13(7)$,
which is close to the expected Onsager value.

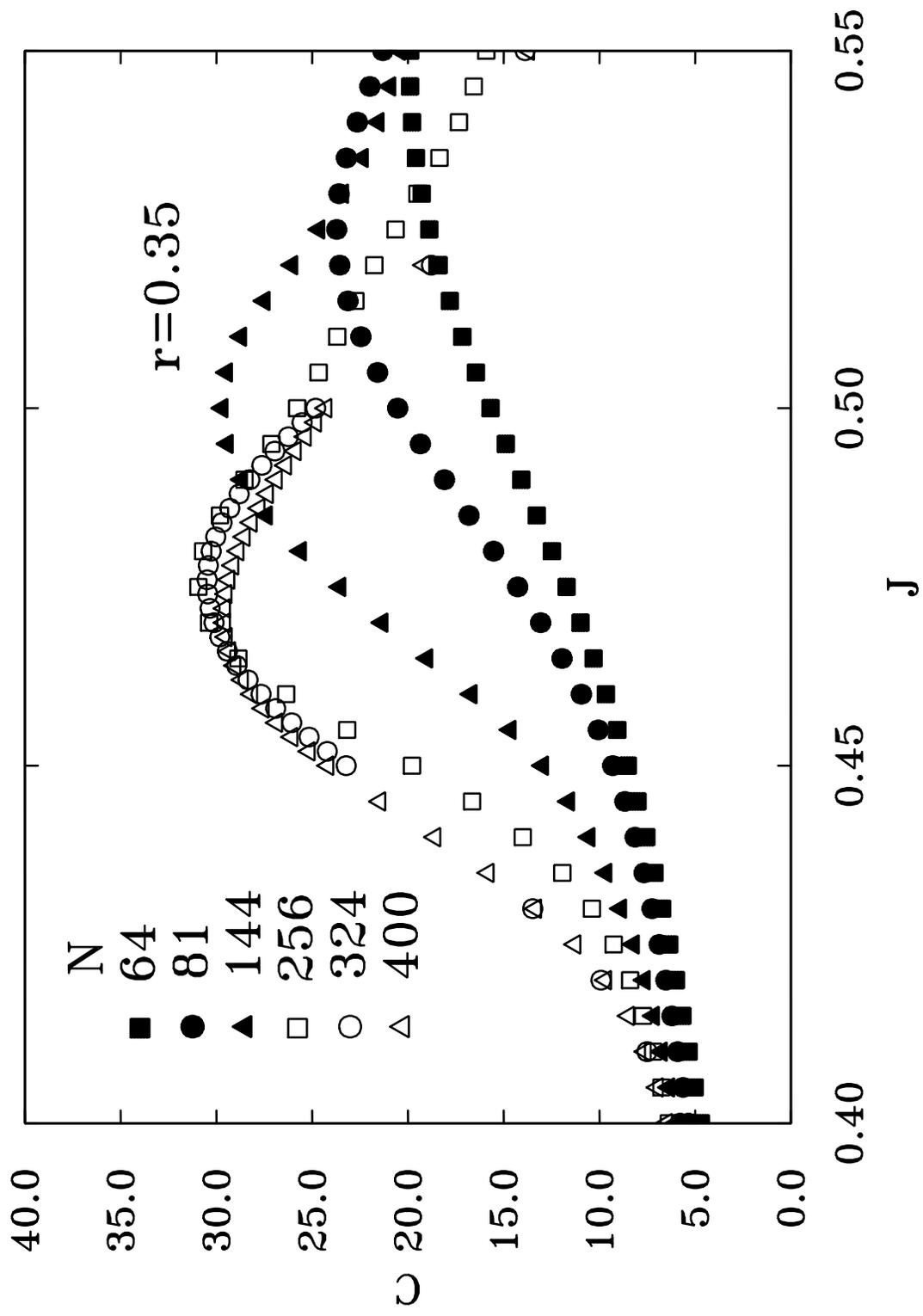
\begin{figure}
\caption{
Plot of the specific heat $C$ versus ferromagnetic coupling $J$ at
$r$=0.35,  showing the absence of a growth in the peak with increasing
lattice size (for the larger systems),
in contrast to the behavior of the magnetic
susceptibility. The errors (not shown) are smaller than the size of the
symbols.
}
\label{heatN}
\end{figure}
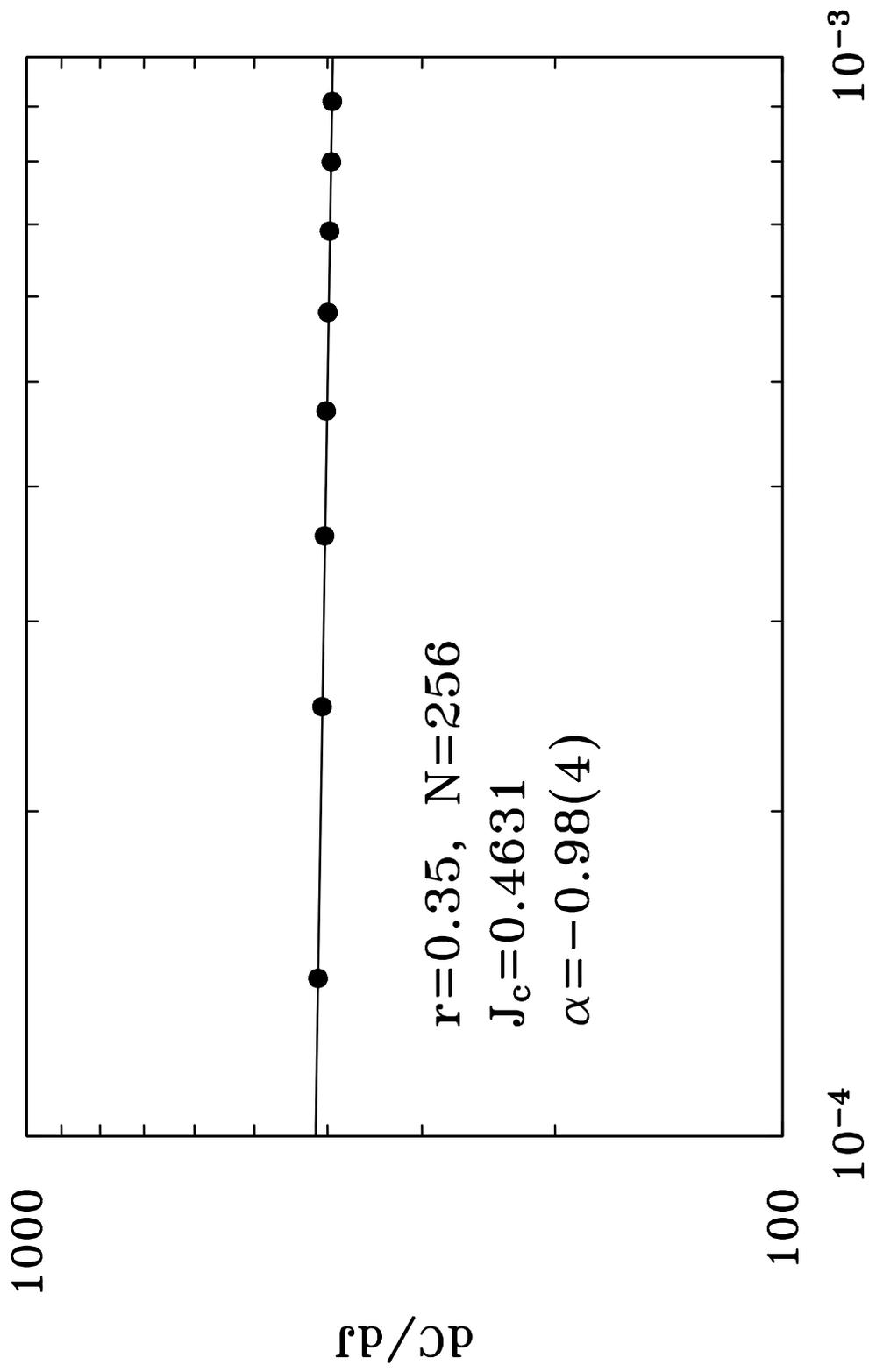
\begin{figure}
\caption{
The derivative of the specific heat $dC/dJ$ as a function of $J_c-J$ on
logarithmic axes for $N=256$.
}
\label{heatJ}
\end{figure}

The results for the Ising specific heat $C$ at the
tricritical point as a function of lattice size $N$ are shown in
Fig.\ \ref{heatN}.
One expects the peak to grow as $C \sim N^{\alpha / 2 \nu} $, but the
absence of any growth for the larger values of $N$
implies that $\alpha / \nu < 0$ (a weak cusp
in the specific heat).
In general close to a critical
point, the free energy can be decomposed into a regular and a singular
part. In our case the singular part  does not seem to be singular
enough to emerge above the regular background, leading to an intrinsic
uncertainty in the determination of an $\alpha < 0$, and which can only
be
overcome by determining still higher derivatives of the free energy
with respect to the coupling $J$. In order to isolate the singular
part of the specific heat we have therefore calculated $dC/dJ$ from the
expression
\beq
{{dC}\over {dJ}} = N^2\left[ 3\langle E \rangle \langle E^2\rangle
-\langle E^3 \rangle - 2 \langle E \rangle^3\right].
\eeq
In the infinite system $dC/dJ$ should diverge according to
\beq
{{dC}\over {dJ}} \sim |J-J_c|^{-(\alpha+1)}.
\label{dCdJ}
\eeq
In particular, if $\alpha=-1$, $dC/dJ$ should diverge logarithmically.
In Fig.\ \ref{heatJ} we show the scaling of
$dC/dJ$ on a lattice with $N=256$ spins according to Eq.~(\ref{dCdJ}). From
the slope of the curve we determine the critical exponent to be
$\alpha\approx -0.98(4)$. We have also tried to assume a logarithmic
scaling behavior as shown in Fig.\ \ref{heatlnJ}. It is clear that from
the linear behavior of $dC/dJ$ we can conclude that our results are
completely consistent with an exponent of $\alpha=-1$. We attribute the
small discrepancy between the results of Figs.\ \ref{heatJ} and \
\ref{heatlnJ} to the fact that we are not sufficiently close to $J_c$
and
that we are on a finite lattice with $N$ sites.
We have also performed a similar analysis for the fluctuation
in the energy per bond (as opposed  to the energy per site as
defined previously). In this case we find close to the tricritical
point $\alpha\approx -0.96(2)$.

In the {\it regular} Ising case one has in a finite volume a logarithmic
divergence $C \sim \ln N $ (and $\alpha / 2 \nu =0$), and we indeed see
such a divergence clearly for $r=0.6$, which corresponds
to the almost regular triangular Ising case.

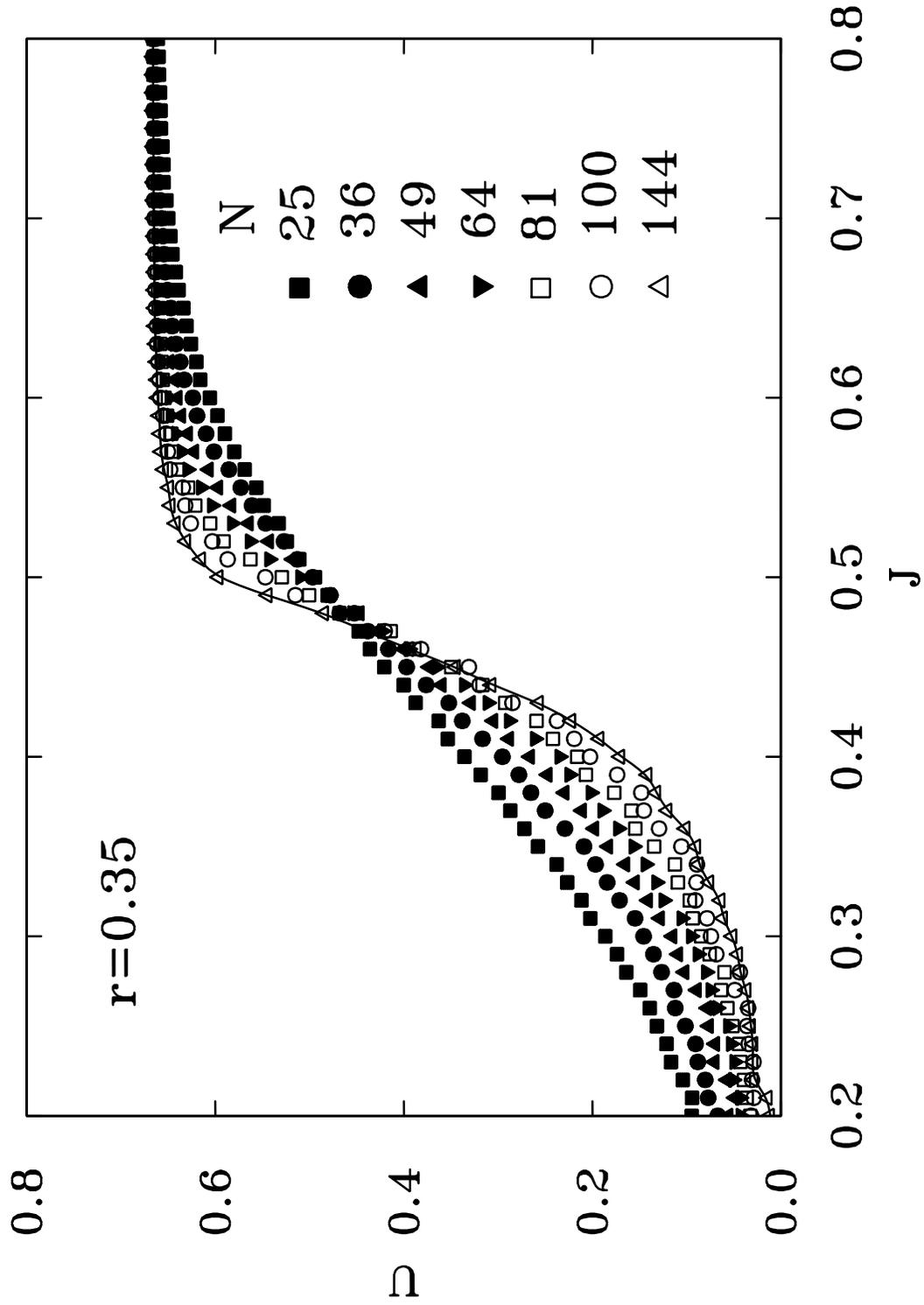
\begin{figure}
\caption{
The Binder fourth-order cumulant $U$ as a function of $J$ for
fixed hard-core repulsion $r=0.35$ and on several lattices with $N$ spins.
The solid line is a spline through the data for $N=144$.
}
\label{binderU}
\end{figure}

Another approach to obtaining $\alpha$ is to determine the correlation
length exponent
$\nu$ directly instead, and use scaling to relate it to $\alpha = 2 - 2 \nu $.
The exponent $\nu$ can be obtained in the following way.
First one can improve on the estimate for $J_c$ by considering the
fourth-order cumulant \cite{binder}
\beq
U_N (J) = 1 - { < m^4 > \over 3 < m^2 >^2 } \;\;\;\; ,
\eeq
where $m = \sum_i S_i / N $.
It has the scaling form expected of a dimensionless quantity
\beq
U_N (J) = \bar U ( N^{1 / 2 \nu} |J -J_c| ) .
\eeq
The curves $ U_N (J) $, for different and sufficiently large values of
$N$,
should then intersect at a common point $J_c$, where
the theory exhibits scale invariance, and $U$ takes on the fixed
point value $U^*$. In Fig.\ \ref{binderU} we show the fourth-order
cumulant as a function of $J$ for $r=0.35$ and for several lattice sizes.
We have found that indeed the curves meet very close to a common point,
and from the intersection of the curves for $N$ = 25 to 400 we estimate
$J_c = 0.472(9)$,
which is consistent with the estimate of the critical point derived
from the location of the peak in the magnetic susceptibility.
We also determine $U^* = 0.47(4)$, which should be compared to
the pure Ising model
estimate for the invariant charge $U^* \approx 0.613$ \cite{ustar}.

One can then estimate the correlation length exponent $\nu$
from the scaling of the slope of the cumulant at $J_c$. For
two lattice sizes $N,N'$ one computes the estimator
\beq
\nu_{eff} (N,N') \; = \; { \ln [ N' / N ] \over
2 \ln [ U'_{N'}(J_c) /  U'_{N}(J_c) ] } \;\;\;\; ,
\eeq
with $ U'_{N} \equiv \partial U_N / \partial J $ defined by
\beq
U'_{N}=
{N\over 3\langle m^2\rangle^2}\left[\langle m^4\rangle \langle E
\rangle + \langle m^4 E \rangle -2{{\langle m^4\rangle \langle m^2 E
\rangle }\over{\langle m^2\rangle}}\right].
\label{dUdJ}
\eeq
Using values of $N$ from systems with 256, 400, and 900 spins we
estimate $\nu$ from Eq.~(\ref{dUdJ}) to be $1.46(8)$. Using the
scaling relationship $\alpha=2-2\nu$, we obtain an estimate for $\alpha$
which is again quite consistent with our previous estimate derived
from $dC/dJ$.

In Table\ \ref{TabIsing} we summarize our results, together with the
exponents
obtained for the two-matrix model \cite{kazakov}
(and which are the same as the KPZ values \cite{kpz}), for the Onsager
solution of the square lattice Ising model, and for
the tricritical Ising model in two dimensions \cite{tricr}.
As can be seen, the exponents are quite close to the matrix model
values (the pure Ising exponents seem to be excluded by several
standard deviations).

\begin{table}
\caption{ Estimates of the critical exponents for the random two-dimensional
Ising model, as obtained from finite-size scaling at the tricritical
point.}
\begin{center}
\begin{tabular}{|l|l|l|l|l|l|}
\hline
& $ \gamma / \nu $ & $\beta / \nu $ & $ \alpha / \nu $ & $ \alpha $ & $ \nu $
\\
\hline \hline
This work    & 1.32(3) & 0.31(4)  & -0.65(4) & -0.98(4) & 1.46(8) \\
\hline
Matrix model & 1.333...  & 0.333... & -0.666... & -1.0 & 1.5  \\ \hline
Onsager      & 1.75    & 0.125  & 0.0 & 0.0 & 1.0  \\ \hline
Tricritical Ising & 1.85  & 0.075 & 1.60 & 0.888... & 0.555... \\ \hline
\end{tabular}
\end{center}
\label{TabIsing}
\end{table}
\vskip 10pt

\section{Conclusions}

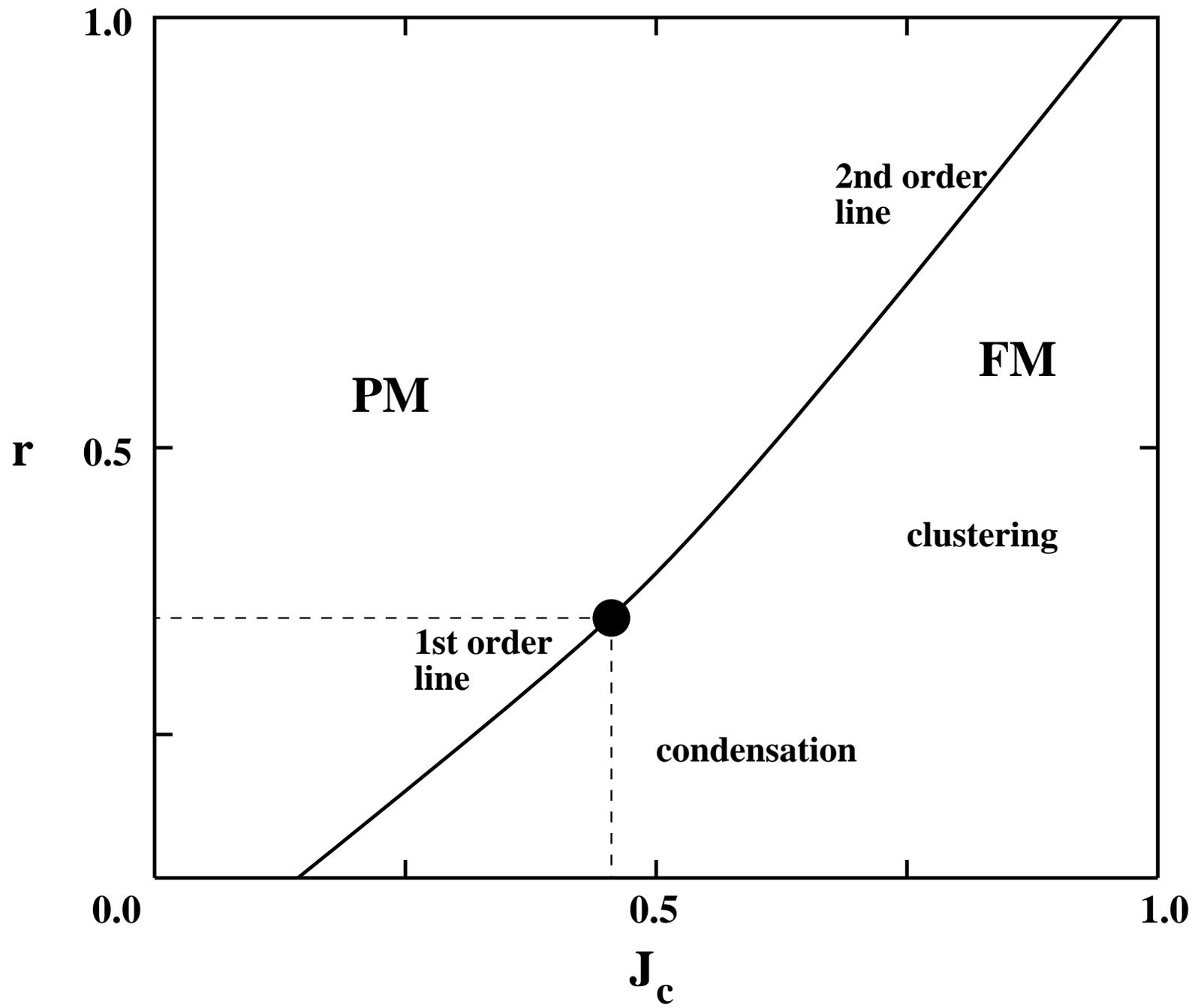
\begin{figure}
\caption{
The phase diagram for the dynamical random Ising model on a
two-dimensional flat surface.
The tricritical point (denoted by the solid circle)
separates the first order from the second order transition lines. The
paramagnetic (PM) and ferromagnetic (FM) phases are also shown.
}
\label{Isingdiag}
\end{figure}

In the previous sections we have presented some results for the
exponents of a
random Ising model in flat two-dimensional space. The model reproduces
some of the features of a model for dynamically triangulated Ising
spins,  and
in particular its random nature, but does not incorporate any effects
due to curvature.
Due to the non-local nature of the interactions of the spins, only
relatively small systems have been considered so far, which is reflected
in the still rather large uncertainties associated with the exponents.
Still a rich phase diagram has emerged, with a tricritical point
separating first from second order transition lines. The phase diagram
we obtain is shown in Fig.\ \ref{Isingdiag}.
We have localized the tricritical point at $J_c = 0.471(5)$ and
$r=0.344(7)$.
The thermal and magnetic exponents determined in the vicinity of
the tricritical point (presented in Table\ \ref{TabIsing}) have been
found to be
consistent, within errors, with the matrix model solution of the random
Ising model and the KPZ values.
Our results would therefore suggest that matrix model solutions can
also be used to describe a class of annealed random systems in flat
space.

One might wonder at this point if the spin system discussed in this paper
can be found among the models in the FQS classification scheme
\cite{fqs} of two-dimensional conformally invariant field theories
\footnote{We thank Giorgio Parisi for suggesting to look into
this aspect.}.
Since the model is apparently not unitary
(it contains short range repulsion and long range
attraction terms), it should fall into the wider class
of degenerate theories considered by BPZ \cite{bpz}.
The allowed scaling dimensions in these theories are given by the
well-known Kac formula,
\begin{equation}
\Delta_{p,q} = { 1 \over 4 } \left [ ( p \alpha_+ + q \alpha_- )^2 -
( \alpha_+ + \alpha_- )^2  \right ]
\end{equation}
with $p,q$ positive integers, and
$ \alpha_{\pm} = \alpha_0 \pm (1 + \alpha_0^2 )^{1/2}$.
$\alpha_0$ is related to the conformal anomaly $c$ of the theory
by $c = 1 - 24 \alpha_0^2 $.
Often the central charge is then written as $c= 1 - 6 / m(m+1)$.
One of the difficulties in this approach is the identification of
a given realization of conformal symmetry with a particular
universality class.
The simplest possibility appears to be $m=4/5$, corresponding to
$m = r / (s-r)$ with $s=9$ and $r=4$. One then obtains for this choice
the central charge $ c = -19/6 $, and
$\alpha_0 = 5/12$, $\alpha_+ = 3/2$ and $\alpha_- = -2/3$.
The matching scaling dimensions are then
$\Delta_{1,4} = \Delta_{3,5} = 1/6 $ (which gives $\eta = 2/3$), and
$\Delta_{1,5} = \Delta_{3,4} = 2/3 $ (which gives $\nu  = 3/2$).
Negative values of $c$ are allowed in non-unitary theories.
It would be of interest to compute the central charge directly in the
random spin model and
verify this assignment, using the methods described in
Ref. \cite{itz}.

We should mention in closing that
the above values for $s$ and $r$ appear to be rather
close to the ones associated
with the Yang-Lee edge singularity, which describes the behavior
of the magnetization in the Ising model in the presence
of an imaginary external field, and for which Cardy \cite{cardy}
has suggested
the identification $s=5$ and $r=2$, which yields $m=2/3$ and
$c=-22/5$. It is known that the Yang-Lee edge singularity
also describes the critical properties of large branched dilute
polymers and of the Ising model in a quenched random external
field in $d+2$ dimension \cite{poly}.

\vspace{12pt}

{\bf Acknowledgements}

One of us (H.W.H.) acknowledges useful discussions
with G. Parisi
during a visit at the University of Rome.
We would also like to thank Martin Hasenbush for his comments
related to this work.
The numerical computations  were performed on
facilities provided by the National Center for Supercomputer
Applications
(NCSA), the San Diego Supercomputer Center (SDSC),
the University of California at Irvine, and by the Texas National Research
Laboratory Commission through grants RGFY9166 and RGFY9266.

\newpage

\vspace{12pt}

\begin{figure}
\caption{
A particular configuration of the system with $N=400$ spins and $r=0.4$
for
$J=0.35$. Spins with $S=\pm 1$ are indicated with empty and solid
circles,
respectively.
}
\label{config2}
\end{figure}

\begin{figure}
\caption{
A particular configuration of the system with $N=400$ spins and $r=0.4$
for
$J=0.65$. Spins with $S=\pm 1$ are indicated with empty and solid
circles,
respectively.
}
\label{config1}
\end{figure}

\begin{figure}
\caption{
A particular configuration of the system with $N=400$ spins and $r=0.98$
for
$J=0.25$. The hard core repulsion radius is shown as a circle around the
spin.
Spins with $S=\pm 1$ are indicated with empty and solid circles,
respectively.
}
\label{config3}
\end{figure}

\begin{figure}
\caption{
The average number of neighbors $z$ as a function of $J$ on a lattice with
$N=144$ spins for several choices of the hard core repulsion $r$.
}
\label{neighbors}
\end{figure}

\begin{figure}
\caption{
The average energy per bond $E_z$ as a function of $J$ for several choices of
the hard core
repulsion $r$ for a system with $N=100$ sites.
}
\label{energy}
\end{figure}

\begin{figure}
\caption{
The derivative of the specific heat $dC/dJ$ as a function of $J_c-J$ on
semi-logarithmic axes for $N=256$.
}
\label{heatlnJ}
\end{figure}

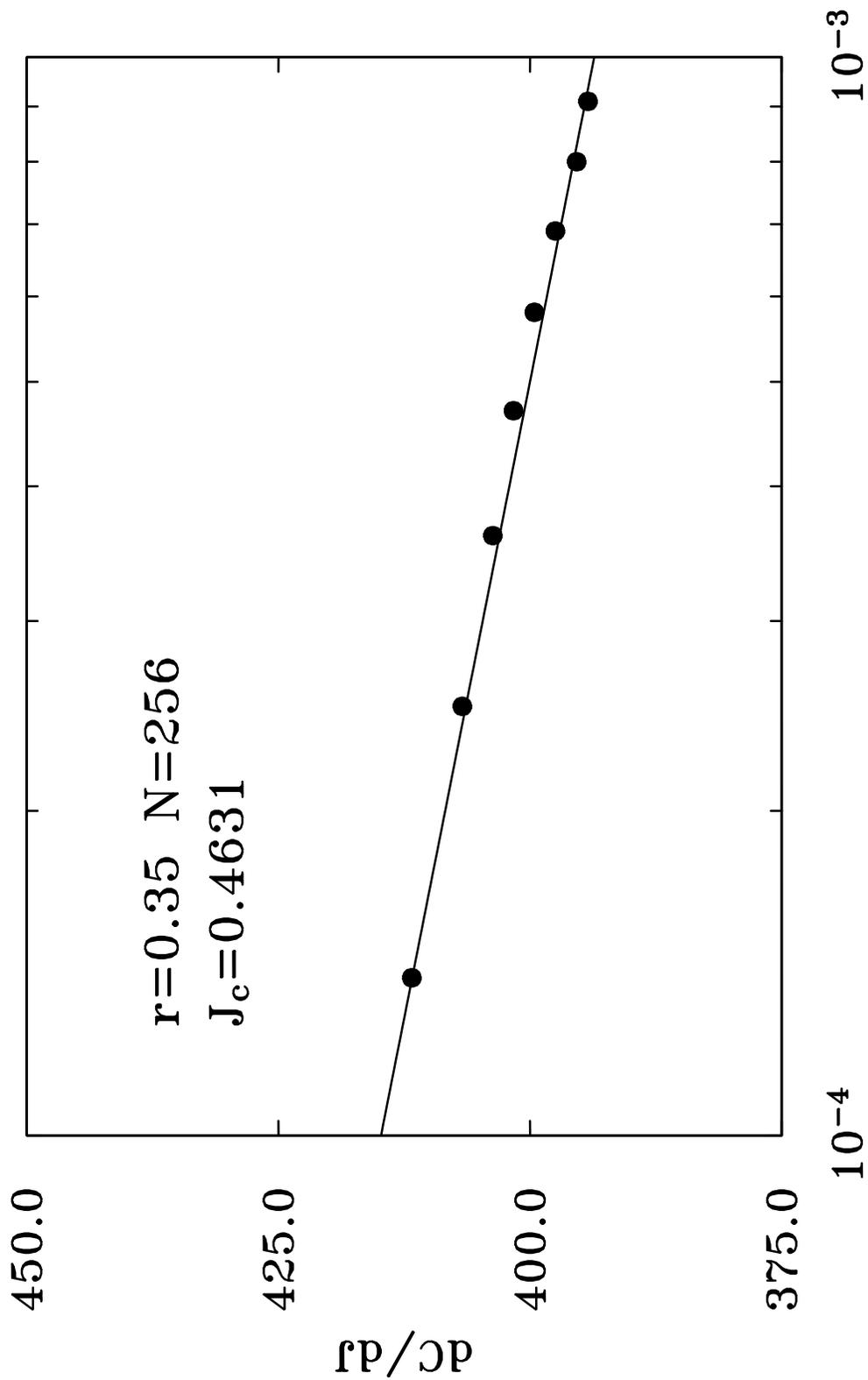
\end{document}